\documentclass[aps,prd,twocolumn,superscriptaddress,showpacs]{revtex4}

\usepackage[dvips]{graphicx}
\usepackage{ulem}
\usepackage{float}
\usepackage[usenames]{color}
\usepackage{amsmath}
\usepackage{hyperref}
\usepackage[all]{hypcap}
\usepackage{float}

\begin{document}

\title{Identification of the newly observed $\Sigma_b(6097)^\pm$ baryons from their strong decays}

\author{Pei Yang}

%\affiliation{Department of Physics, Shanghai University, Shanghai 200444, China}

\author{Jing-Jing Guo}

%\affiliation{Department of Physics, Shanghai University, Shanghai 200444, China}

%\author{Hao-Yang Jing}

%\affiliation{Department of Physics, Shanghai University, Shanghai 200444, China}

\author{Ailin Zhang}

\email{zhangal@staff.shu.edu.cn}

\affiliation{Department of Physics, Shanghai University, Shanghai 200444, China}

\begin{abstract}
Two bottom $\Sigma_b(6097)^\pm$ baryons were observed in the final states $\Lambda_b^0\pi^-$ and $\Lambda_b^0\pi^+$ in $pp$ collision by LHCb collaboration, whose masses and widths were measured. In a $^{3}P_{0}$ model, the strong decay widths of two ground $S$-wave and seven excited $P$-wave $\Sigma_b$ baryons have been systematically computed. Numerical results indicate that the newly observed $\Sigma_b(6097)^\pm$ are very possibly $\Sigma_{b2}^1({3\over 2}^-)$ with $J^P={3\over 2}^-$ or $\Sigma_{b2}^1({5\over 2}^-)$ with $J^P={5\over 2}^-$. The predicted decay widths of $\Sigma_b(6097)^\pm$ are consistent with experimental measurement from LHCb. In particular, it may be possible to distinguish these two assignments through ratios $\Gamma({\Sigma_b(6097)^\pm\to \Sigma_b^\pm\pi^0})/\Gamma({\Sigma_b(6097)^\pm\to \Sigma_b^{*\pm}\pi^0})$, which can be measured by experiments in the future. In the meantime, our results support the assignments that $\Sigma_b^\pm$ and $\Sigma_b^{*\pm}$ are the ground $S$-wave $\Sigma_b$ baryons with $J^P={1\over 2}^+$ and $J^P={3\over 2}^+$, respectively.
\end{abstract}


\maketitle

\section{Introduction \label{sec:introduction}}
There are two light u, d quarks and one heavy b quark in $\Sigma_b$ baryons, and the two light quarks couple to isospin $1$ inside. Four $\Sigma_{b}^\pm$ and $\Sigma_{b}^{*\pm}$ have been observed by the CDF collaboration~\cite{cdf,cdf2}. Their spins or parities have not been measured by experiment, they are assigned as the ground $S$-wave $\Sigma_b$ with $J^P={1\over 2}^+$ and $J^P={3\over 2}^+$, respectively, in quark models. The assignments need confirmation in more ways. The masses and widths of these baryons from Particle Data Group~\cite{pdg} are given in Table~\ref{tab:spectrum}. Recently, the data were precisely improved by LHCb experiment~\cite{lhcb}.

\begin{table*}[t]
	%\large
\begin{center}
\caption{Masses and widths of the ground $\Sigma_b$ (in MeV)~\cite{pdg}.}\label{tab:spectrum}
\begin{tabular}{l  cccc  c  c  c  c }\hline \hline
~State~~~~~~ & ~~~~~~~~$J^P$~~~~~~~~ &
~~~~~~~~~Mass~~~~~~~~~ & ~~~~~Width~~~~~ &~~~~~Decay modes~~~~~\\
\hline
$\Sigma_b^{+}$ & ${1\over 2}^+$ & $5811.3\pm1.3$ & $9.7^{+4.0}_{-3.0}$ &
$\Lambda_b^{0}\pi$\\

$\Sigma_b^{-}$ &${1\over 2}^+$&  $5815.5\pm1.8$ & $4.9^{+3.3}_{-2.4}$ &
$\Lambda_b^{0}\pi$\\

$\Sigma_b^{*+}$ & ${3\over 2}^+$ &  $5832.1\pm1.9$ & $11.5\pm2.8$ &
$\Lambda_b^{0}\pi$\\

$\Sigma_b^{*-}$ & ${3\over 2}^+$ &  $5835.1\pm1.9$ & $7.5\pm2.3$ &
$\Lambda_b^{0}\pi$\\
\hline\hline
										
\end{tabular}
\end{center}
\end{table*}	

In the same LHCb experiment, two bottom $\Sigma_b(6097)^\pm$ baryons were first observed in final states $\Lambda_b^0\pi^-$ and $\Lambda_b^0\pi^+$ in $pp$ collision. The masses and widths of the $\Sigma_b(6097)^\pm$ are measured
\begin{eqnarray*}
m(\Sigma_{b}(6097)^-)=6098.0\pm~1.7\pm~0.5~\rm{MeV}, \\
\Gamma(\Sigma_{b}(6097)^-)=28.9\pm~4.2\pm~0.4~\rm{MeV}, \\
m(\Sigma_{b}(6097)^+)=6095.8\pm~1.7\pm~0.4~\rm{MeV}, \\
\Gamma(\Sigma_{b}(6097)^+)=31.0\pm~5.5\pm~0.7~\rm{MeV}. \\
\end{eqnarray*}

The identification of heavy baryons provides an excellent way to explore the structure and dynamics in baryons~\cite{capstick,roberts,klempt,crede,cheng}. Therefore, the identification of $\Sigma_b(6097)^\pm$ is an important topic in the quark model. In Ref.~\cite{chen1}, $\Sigma_b(6097)^\pm$ were explained as $P$-wave baryons with $J^P={3\over 2}^-$ or $J^P={5\over 2}^-$ based on the mass spectrum analysis and the strong decay calculation in a diquark picture. In Ref.~\cite{zhong1}, $\Sigma_b(6097)^\pm$ were also explained as $P$-wave baryons with $J^P={3\over 2}^-$ or $J^P={5\over 2}^-$ based on their strong decay analysis in a chiral quark model.

As a phenomenological method, $^3P_0$ model has been employed to compute the OZI-allowed hadronic decay widths of hadrons after its appearance~\cite{micu1969,yaouanc1,yaouanc2,yaouanc3}. Though the bridge between the phenomenological $^3P_0$ model and QCD has not been established, some attempts have been made~\cite{swanson,ackleh,bonnaz}. The $^3P_0$ model is also capable of exploring the dynamics and structure of baryons or multi-quark systems. Recently, the approach has been employed to study of the structure of charmed baryons through their strong decays~\cite{zhu,zhang,zhang2,zhang3,zhang4}. In this work, we will study the $P$-wave possibility of $\Sigma_b(6097)^\pm$ in detail. By the way, the ground $S$-wave $\Sigma_b$ possibility of $\Sigma_b$ and $\Sigma_b^*$ will be examined.

The work is organized as follows. In Sec.II, the $^3P_0$ model is briefly introduced, some notations of heavy baryons and related parameters are indicated. We present our numerical results and analyses in Sec.III. In the last section, we give our conclusions and discussions.

\section{$^3P_0$ model, some notations and parameters\label{Sec: $^3P_0$ model}}
$^3P_0$ model is also called a Quark Pair Creation (QPC) model. It was first proposed by Micu\cite{micu1969} and further developed by Yaouanc {\it et al}~\cite{yaouanc1,yaouanc2,yaouanc3}. The basic idea of this model is assumed that a pair of quark $q\bar{q}$ is created from the QCD vacuum
with vacuum quantum numbers $J^{PC}=0^{++}$, and then regroup with the quarks from the initial hadron A to form two daughter hadrons B and C~\cite{micu1969}. For a bottom baryon, there are three ways of regrouping shown in the following equations~\cite{zhu,zhang4}:
\begin{eqnarray}
&{\mathcal{A}}(u_1, u_2, b_3)+\mathcal{P}(u_4,
\bar{u}_5)\rightarrow{\mathcal{B}}(u_1, u_2,
u_4)+{\mathcal{C}}(b_3, \bar{u}_5),{ }\label{com-1}\\
&{\mathcal{A}}(u_1, u_2, b_3)+\mathcal{P}(u_4,
\bar{u}_5)\rightarrow{\mathcal{B}}(u_1, b_3,
u_4)+{\mathcal{C}}(u_2, \bar{u}_5),{ }\label{com-2}\\
&{\mathcal{A}}(u_1, u_2, b_3)+\mathcal{P}(u_4,
\bar{u}_5)\rightarrow{\mathcal{B}}(u_2, b_3,
u_4)+{\mathcal{C}}(u_1, \bar{u}_5),{ }\label{com-3}
\end{eqnarray}
where $uub$ (the constituent quark of initial baryon A) could be replaced by $ddb$, and the created quark pair $u\bar{u}$ could be replaced by $d\bar{d}$.

In the $^3P_0$ model, the hadronic decay width $\Gamma$ of a process $A \to B + C$ follows~\cite{yaouanc3},
\begin{eqnarray}
\Gamma  = \pi ^2 \frac{|\vec{p}|}{m_A^2} \frac{1}{2J_A+1}\sum_{M_{J_A}M_{J_B}M_{J_C}} |{\mathcal{M}^{M_{J_A}M_{J_B}M_{J_C}}}|^2.
\end{eqnarray}
In the equation, $\vec{p}$ is the momentum of the daughter baryon in A's center of mass frame,
\begin{eqnarray}
 |\vec{p}|=\frac{{\sqrt {[m_A^2-(m_B-m_C )^2][m_A^2-(m_B+m_C)^2]}}}{{2m_A }}.
\end{eqnarray}
$m_A$ and $J_A$ are the mass and the total angular momentum of the initial baryon A, respectively. $m_B$ and $m_C$ are the masses of the final hadrons. $\mathcal{M}^{M_{J_A}M_{J_B}M_{J_C}}$ is the helicity amplitude, which reads~\cite{zhu,zhang,zhang2,zhang4}
\begin{flalign}
 &\delta ^3 (\vec{p}_B + \vec{p}_C -\vec{p}_A) \mathcal{M}^{M_{J_A } M_{J_B } M_{J_C }}\nonumber \\
 &=-2\gamma\sqrt {8E_A E_B E_C }  \sum_{M_{\rho_A}}\sum_{M_{L_A}}\sum_{M_{\rho_B}}\sum_{M_{L_B}} \sum_{M_{S_1},M_{S_3},M_{S_4},m}  \nonumber\\
 &\langle {J_{l_A} M_{J_{l_A} } S_3 M_{S_3 } }| {J_A M_{J_A } }\rangle \langle {L_{\rho_A} M_{L_{\rho_A} } L_{\lambda_A} M_{L_{\lambda_A} } }| {L_A M_{L_A } }\rangle \nonumber \\
 &\langle L_A M_{L_A } S_{12} M_{S_{12} }|J_{L_A} M_{J_{L_A} } \rangle \langle S_1 M_{S_1 } S_2 M_{S_2 }|S_{12} M_{S_{12} }\rangle \nonumber \\
 &\langle {J_{l_B} M_{J_{l_B} } S_3 M_{S_3 } }| {J_B M_{J_B } }\rangle \langle {L_{\rho_B} M_{L_{\rho_B} } L_{\lambda_B} M_{L_{\lambda_B} } }| {L_B M_{L_B } }\rangle \nonumber \\
 &\langle L_B M_{L_B } S_{14} M_{S_{14} }|J_{l_B} M_{J_{J_B} } \rangle \langle S_1 M_{S_1 } S_4 M_{S_4 }|S_{14} M_{S_{14} }\rangle \nonumber \\
 &\langle {1m;1 - m}|{00} \rangle \langle S_4 M_{S_4 } S_5 M_{S_5 }|1 -m \rangle \nonumber \\
 &\langle L_C M_{L_C } S_C M_{S_C}|J_C M_{J_C} \rangle \langle S_2 M_{S_2 } S_5 M_{S_5 }|S_C M_{S_C} \rangle \nonumber \\
&\times\langle\varphi _B^{1,4,3} \varphi _C^{2,5}|\varphi _A^{1,2,3}\varphi _0^{4,5} \rangle \times I_{M_{L_B } ,M_{L_C } }^{M_{L_A },m} (\vec{p}).
\end{flalign}
The factor $2$ in front of $\gamma$ results from the fact that Eq.~(\ref{com-2}) and Eq.~(\ref{com-3}) give the same final states.

In the equation above, the matrix $\langle \varphi_B^{1,4,3} \varphi_C^{2,5}|\varphi_A^{1,2,3}\varphi_0^{4,5} \rangle$ of the flavor wave functions $\varphi_{i}$ $(i=A,B,C,0)$ can also be computed in terms of a matrix of the isospins as follows~\cite{yaouanc3,zhang}
\begin{flalign}
%\scriptstyle
\langle\varphi_B^{1,4,3} \varphi_C^{2,5}|\varphi_A^{1,2,3}\varphi_0^{4,5} \rangle =\mathcal{F}^{(I_A;I_BI_C)}<I_BI_B^3I_CI_C^3|I_AI_A^3>
\end{flalign}
with
\begin{flalign}
\mathcal{F}^{(I_A;I_BI_C)}&=f \cdot (-1)^{I_{12}+I_C+I_A+I_3}   \nonumber \\
&\times [\frac{1}{2}(2I_C  + 1)(2I_B  + 1)]^{1/2} \nonumber \\
&\times \begin{Bmatrix}
    {I_{12}} & {I_B} & {I_4}\\
    {I_C} & {I_3} & {I_A}\\ \end{Bmatrix}
\end{flalign}
where $f$ takes the value of $(\frac{2}{3})^{1/2}$ or $-(\frac{1}{3})^{1/2}$ for the isospin $\frac{1}{2}$ or $0$ of the created quarks, respectively. $I_{A}$, $I_B$ and $I_M$ represent the isospins of the initial baryon, the final baryon and the final meson. $I_{12}$, $I_{3}$ and $I_{4}$ denote the isospins of relevant quarks, respectively.

The space integral follows as,
\begin{flalign}
I_{M_{L_B } ,M_{L_C } }^{M_{L_A } ,m} (\vec{p})&= \int d \vec{p}_1 d \vec{p}_2 d \vec{p}_3 d \vec{p}_4 d \vec{p}_5 \nonumber \\
&\times\delta ^3 (\vec{p}_1 + \vec{p}_2 + \vec{p}_3 -\vec{p}_A)\delta ^3 (\vec{p}_4+ \vec{p}_5)\nonumber \\
&\times \delta ^3 (\vec{p}_1 + \vec{p}_4 + \vec{p}_3 -\vec{p}_B )\delta ^3 (\vec{p}_2 + \vec{p}_5 -\vec{p}_C) \nonumber \\
& \times\Psi _{B}^* (\vec{p}_1, \vec{p}_4,\vec{p}_3)\Psi _{C}^* (\vec{p}_2 ,\vec{p}_5) \nonumber \\
& \times \Psi _{A} (\vec{p}_1 ,\vec{p}_2 ,\vec{p}_3)y _{1m}\left(\frac{\vec{p_4}-\vec{p}_5}{2}\right)
\end{flalign}

with a simple harmonic oscillator (SHO) wave functions for the baryons~\cite{capstick2,roberts2,zhang}
\begin{flalign}
\Psi_{}(\vec{p}_{})&=N\Psi_{n_{\rho} L_{\rho} M_{L_{\rho}}}(\vec{p}_{\rho}) \Psi_{n_{\lambda} L_{\lambda} M_{L_{\lambda}}}(\vec{p}_{\lambda}),
\end{flalign}
where $N=3^\frac{3}{4}$ represents a normalization coefficient of the total wave function. Explicitly,
\begin{flalign}
\Psi_{nLM_L}(\vec{p})&=\frac{(-1)^n(-i)^L}{\beta^{3/2}}\sqrt{\frac{2n!}{\Gamma(n+L+\frac{3}{2})}}\big(\frac{\vec{p}}{\beta}\big)^L \exp(-\frac{\vec{p}^2}{2\beta^2}) \nonumber\\
&\times L_n^{L+1/2}\big(\frac{\vec{p}^2}{\beta^2}\big)Y_{LM_L}(\Omega_p)
\end{flalign}
where $L_n^{L+1/2}\big(\frac{\vec{p}^2}{\beta^2}\big)$ denotes the Laguerre polynomial function, and $Y_{LM_L}(\Omega_p)$ is a spherical harmonic function. The relation between the solid harmonica polynomial $y _{1m}(\vec{p})$ and $Y_{LM_L}(\Omega_{\vec{p}})$ is $y _{1m}(\vec{p})=|\vec{p}|^L Y_{LM_L}(\Omega_p)$. In above equations, Jacobi coordinates $\vec{\rho}$ and $\vec{\lambda}$~\cite{jacobi} were employed.

Notations and internal structures of the heavy baryons in quark model are explained in Refs.~\cite{cheng2,zhu,zhang3,zhang4}. In this model, there are two $S$-wave and seven $P$-wave $\Sigma_b$. In other quark model, the structure and dynamics in baryons may be different. In fact, the difference is an indication of the complexity of baryon. Accordingly, the numbers of $P$-wave $\Sigma_b$ may be different in different models. For example, there are five $P$-wave $\Sigma_b$ in Refs.~\cite{ebert,rosner,rai,chen1,zhong1}. For a practical calculation, the quantum numbers of two $1S$-wave and seven $1P$-wave $\Sigma_b$ baryons are presented in Table~\ref{table4}.

\begin{table}[h]
\caption{Quantum numbers of $1S$-wave and $1P$-wave excitations of $\Sigma_{b}$}
\begin{tabular}{lc c cccccccc}
\hline\hline
&N	& ~~Assignments~~  & ~$J$ ~ & ~$J_l$ ~& ~$L_\rho$ ~& ~$L_\lambda$ ~&~ $L$ ~ &~ $S_\rho$ \\
\hline
&1   &$\Sigma_{b1}^{0}(\frac{1}{2}^+)$    & $\frac{1}{2}$ &  1  &  0    &   0    &  0   &  1       \\
&2  &$\Sigma_{b1}^{0}(\frac{3}{2}^+)$     & $\frac{3}{2}$ &  1  &  0    &   0    &  0   &  1       \\
\hline
&3  &$\Sigma_{b0}^{1 }(\frac{1}{2}^-)$    & $\frac{1}{2}$ &  0  &  0    &   1    &  1   &  1       \\
&4  &$\Sigma_{b1}^{1 }(\frac{1}{2}^-)$    & $\frac{1}{2}$ &  1  &  0    &   1    &  1   &  1       \\
&5  &$\Sigma_{b1}^{1 }(\frac{3}{2}^-)$    & $\frac{3}{2}$ &  1  &  0    &   1    &  1   &  1       \\
&6  &$\Sigma_{b2}^{1}(\frac{3}{2}^-)$     & $\frac{3}{2}$ &  2  &  0    &   1    &  1   &  1       \\
&7  &$\Sigma_{b2}^{1 }(\frac{5}{2}^-)$    & $\frac{5}{2}$ &  2  &  0    &   1    &  1   &  1       \\
&8  &$\tilde\Sigma_{b1}^{1 }(\frac{1}{2}^-)$   & $\frac{1}{2}$  &  1    &   1    &  0   &  1   &  0       \\
&9  &$\tilde\Sigma_{b1}^{1 }(\frac{3}{2}^-)$   & $\frac{3}{2}$  &  1    &   1    &  0   &  1   &  0       \\
		\hline\hline\label{table4}
	\end{tabular}	
\end{table}

In this table, $L_\rho$ denotes an orbital angular momentum between the two light quarks, $L_\lambda$ denotes the orbital angular momentum between the bottom quark and the two light quark system, and $L$ is the total orbital angular momentum of $L_\rho$ and $L_\lambda$ ($L$ =$L_\rho$ + $L_\lambda$).
$S_\rho$ denotes the total spin of the two light quarks, $J_l$ is the total angular momentum of $L$ and $S_\rho$ ($J_l$ = $L$ + $S_\rho$), and $J$ is the total angular momentum of the baryons ($J$ = $J_l$ + ${\frac{1}{2}}$). For $\tilde\Sigma_{bJ_l}^{\ L}$, a superscript $\ L$ is specialized to denote different total angular momentum. The tilde indicates $L_\rho=1$, and the blank indicates $L_\rho=0$.
	
Masses of relevant mesons and baryons involved in our calculation are presented in Table~\ref{table3}~\cite{pdg,lhcb}.

\begin{table}[h]
	\caption{Masses of mesons and baryons involved in the decays~\cite{pdg}}
	\begin{tabular}{p{0.0cm} p{2.0cm}p{2.0cm}|p{2.0cm}p{2.0cm}}
		\hline\hline
		&State              &Mass (MeV)  & State          &Mass (MeV)\\
		\hline
		&$\pi^{\pm}      $ &139.570        &$\Sigma_b^{-}$         & 5815.64 \\
		&$\pi^{0}        $ &134.977        &$\Sigma_b^{+}$         & 5810.55 \\
	    &$\Lambda_b^{0}  $ &5619.58        &$\Sigma_b^{*-}$        & 5834.73 \\
		&$\Lambda_b^{0}(5912)  $ &5912.2   &$\Sigma_b^{*+}$        & 5830.28 \\
		&$\Lambda_b^{0}(5920)  $ &5919.58  &$\Sigma_{b}^{+}(6097)$ &6095.8   \\
		&                 -&-              &$\Sigma_{b}^{-}(6097)$ &6098.0     \\
		\hline\hline
	\end{tabular}
	\label{table3}
\end{table}
		
Some parameters are chosen as follows~\cite{godfrey,zhu,godfrey2,godfrey3,zhang}. The dimensionless pair-creation strength $\gamma=13.4$, the $\beta_{\lambda,\rho}=600$ MeV in the $S$-wave baryons wave functions and $\beta_{\lambda,\rho}=400$ MeV in the $P$-wave baryons wave functions, and the $R=2.5$ GeV$^{-1}$ in $\pi$ wave functions.

\section{Numerical results\label{Sec: numerical results}}			
\subsection{Decays of $\Sigma_{b}$ and $\Sigma_{b}^*$}

$\Sigma_b^\pm$ and $\Sigma_b^{*\pm}$ were first observed in the final states $\Lambda_b^0\pi^\pm$ in $p\bar{p}$ collision by the CDF collaboration~\cite{cdf}, and were interpreted as the lowest-lying $\Sigma_b^\pm$ and $\Sigma_b^{*\pm}$ baryons with $J^P=\frac{1}{2}^+$ and $\frac{3}{2}^+$, respectively, according to their decay widths and masses. $\Lambda_b^0\pi^+$ is the only decay mode of $\Sigma_{b}^+$ and $\Sigma_{b}^{*+}$, and $\Lambda_b^0\pi^-$ is the only decay mode of $\Sigma_b^-$ and $\Sigma_b^{*-}$. In the $^3P_0$ model, the hadronic decay widths of these four observed $\Sigma_b^\pm$ into $\Lambda_b^0\pi^\pm$ in two $S$-wave and seven $P$-wave assignments are computed and presented in Table~\ref{table21}, where a `0` indicates a vanish decay channel. In comparison with experimental results (see Table ~\ref{tab:spectrum}), our numerical results support the assignments that $\Sigma_{b}^{\pm}$ and $\Sigma_{b}^*$ are $\Sigma_{b1}^{0}(\frac{1}{2}^+)$ and $\Sigma_{b1}^{0}(\frac{3}{2}^+)$, respectively. $\Sigma_b$ and $\Sigma_b^*$ are very possibly the ground $S$-wave $\Sigma_b$, where there is no $\lambda$ or $\rho$ excitation inside.
\begin{center}
\begin{table}[t]
\caption{Decay widths (MeV) of $\Sigma_{b}^{\pm}$ and $\Sigma_{b}^{*\pm}$ into $\Lambda_{b}^{0}\pi^\pm$ as  $1S$-wave states or $1P$-wave excitations. }
\begin{tabular}{lc c|cc|ccccccc}
\hline\hline
&N~&$\Sigma_{bJ_l} (J^P)$&$\Sigma_b^{+}$ &$\Sigma_b^{-}$ &$\Sigma_b^{*+}$ &$\Sigma_b^{*-}$ & \\
\hline\hline
&1&$\Sigma_{b1}^{0}(\frac{1}{2}^+)$ &4.42 &5.32   & 8.49&9.64  \\
&2&$\Sigma_{b1}^{0}(\frac{3}{2}^+)$ &4.42 &5.32  & 8.49&9.64 \\
&3&$\Sigma_{b0}^{1}(\frac{1}{2}^-)$&64.31&69.16 &83.10& 87.33 \\
&4&$\Sigma_{b1}^{1}(\frac{1}{2}^-)$ &0    &  0    &   0  &0     \\
&5&$\Sigma_{b1}^{1}(\frac{3}{2}^-)$ &0    &  0    &   0  &0     \\
&6&$\Sigma_{b2}^{1}(\frac{3}{2}^-)$ &0.01  &  0.01  & 0.03 &0.03  \\
&7&$\Sigma_{b2}^{1}(\frac{5}{2}^-)$ &0.01  &  0.01  & 0.03 &0.03  \\
&8&$\tilde\Sigma_{b1}^{1 }(\frac{1}{2}^-)$ &0  & 0 & 0    &0  \\
&9&$\tilde\Sigma_{b1}^{1 }(\frac{3}{2}^-)$ &0  & 0 & 0   &0   \\
\hline\hline
\end{tabular}
\label{table21}
\end{table}
\end{center}	

\subsection{Decays of $\Sigma_b(6097)$}

As pointed out in the second section, there are seven $P$-wave $\Sigma_b$ baryons. The masses of low-lying bottom baryons have been systemically predicted in many references such as~\cite{ebert,rosner,hosaka,rai}. If $\Sigma_b(6097)^\pm$ are $P$-wave $\Sigma_b$ baryons, there exist five possible OZI-allowed hadronic decay modes. The five channels are : $\Lambda_{b}^{0}\pi^\pm$, $\Sigma_{b}^\pm\pi^0$, $\Sigma_{b}^{*\pm}\pi^0$, $\Lambda_{b}^{0}(5912)\pi^\pm$ and $\Lambda_{b}^{0}(5920)\pi^\pm$. The strong decay widths of $\Sigma_b(6097)^-$ into these five channels are calculated in seven different $P$-wave assignments, and presented in Table~\ref{table22}. The results of $\Sigma_b(6097)^+$ are presented in Table~\ref{table23}.

\begin{center}
	\setlength{\tabcolsep}{4mm}
\begin{table*}[t]
\caption{Decay widths (MeV) of $\Sigma_{b}^{-}(6097)$ as $1P$-wave excitations. $\mathcal{B}=\Gamma(\Sigma_{b}^{-}(6097)\to\Lambda_{b}^{0}\pi^-)/\Gamma_{total}$. }
\begin{tabular}{lc|ccccccc}
		\hline\hline
		N&$\Sigma_{bJ_l} (J^P)$  & $\Lambda_{b}^{0}\pi^-$ & $\Sigma_{b}^{-}\pi^0$& $\Sigma_{b}^{*-}\pi^0$
		  &$\Lambda_{b}^{0}(5912)\pi^-$&$\Lambda_{b}^{0}(5920)\pi^-$ &$\Gamma_{total}$&$\mathcal{B}$ \\\hline
		3 &$\Sigma_{b0}^{1}(\frac{1}{2}^-)$ & 274.69     & 0       & 0        & 3.44     & 4.90    & 283.03  &97.05\%  \\
		4 &$\Sigma_{b1}^{1}(\frac{1}{2}^-)$ & 0          & 122.97  & 0.28     & 0      & 6.93      & 130.18  &0      \\
		5 &$\Sigma_{b1}^{1}(\frac{3}{2}^-)$ & 0          & 0.24    & 110.23   & 4.88     & 3.46    & 118.81  &0        \\
		6 &$\Sigma_{b2}^{1}(\frac{3}{2}^-)$ & 14.56      & 0.43    & 0.26     & 1.39     & 3.26    & 19.90   &73.17\%  \\
		7 &$\Sigma_{b2}^{1}(\frac{5}{2}^-)$ & 14.56      & 0.19    & 0.40     & 2.39     & 2.55    & 20.09   &72.47\%  \\
		8 & $\tilde\Sigma_{b1}^{1 }(\frac{1}{2}^-)$& 0   & 184.46   & 1.73    & 0   & 0  & 186.19  & 0                \\
		9 & $\tilde\Sigma_{b1}^{1 }(\frac{3}{2}^-)$ & 0  & 1.43     & 166.00  & 0   & 0  & 167.43  & 0                \\
		\hline \hline
\end{tabular}
\label{table22}
\end{table*}
\end{center}

\begin{center}
	\setlength{\tabcolsep}{4mm}
\begin{table*}[t]
\caption{Decay widths (MeV) of $\Sigma_{b}^{+}(6097)$ as $1P$-wave excitations. $\mathcal{B}=\Gamma(\Sigma_{b}^{+}(6097)\to\Lambda_{b}^{0}\pi^+)/\Gamma_{total}$.}
		\begin{tabular}{lc|ccccccc}
		\hline\hline
		N&$\Sigma_{bJ_l} (J^P)$  & $\Lambda_{b}^{0}\pi^+$ & $\Sigma_{b}^{+}\pi^0$& $\Sigma_{b}^{*+}\pi^0$ &$\Lambda_{b}^{0}(5912)\pi^+$&$\Lambda_{b}^{0}(5920)\pi^+$ &$\Gamma_{total}$&$\mathcal{B}$\\\hline
		3 & $\Sigma_{b0}^{1}(\frac{1}{2}^-)$ & 275.01    & 0      & 0      & 3.14     & 4.40    & 282.55  &97.33\%    \\
		4 & $\Sigma_{b1}^{1}(\frac{1}{2}^-)$ & 0         & 124.85 & 0.31   & 0        & 6.22    & 131.38  &0          \\
		5 & $\Sigma_{b1}^{1}(\frac{3}{2}^-)$ & 0         &0.26    & 111.80 & 4.45     & 3.11    & 119.62  &0          \\
		6 & $\Sigma_{b2}^{1}(\frac{3}{2}^-)$ & 14.19     & 0.46   & 0.27   & 1.27     & 2.93    & 19.12   &74.22\%    \\
		7 & $\Sigma_{b2}^{1}(\frac{5}{2}^-)$ & 14.19     & 0.21   & 0.43   & 2.18     & 2.29    & 19.30    &73.52\%    \\
		8 & $\tilde\Sigma_{b1}^{1 }(\frac{1}{2}^-)$ & 0   & 187.28   & 1.84       & 0    & 0    & 189.12   &0          \\
		9 &  $\tilde\Sigma_{b1}^{1 }(\frac{3}{2}^-)$& 0   & 1.54     & 168.39     & 0    & 0    & 169.93   &0          \\
		 \hline\hline
\end{tabular}
\label{table23}
\end{table*}
\end{center}

In the calculation, $\Sigma_{b}$ and $\Sigma_{b}^*$ are set to the ground $S$-wave $\Sigma_{b1}^{0}(\frac{1}{2}^+)$ and $\Sigma_{b1}^{0}(\frac{3}{2}^+)$ as indicated in previous subsection. $\Lambda_{b}^{0}(5912)$ and $\Lambda_{b}^{0}(5920)$ were observed in $\Lambda_b^0\pi^+\pi^-$ in $pp$ collision by LHCb~\cite{lhcb2}, and interpreted as the orbitally excited $\Lambda_b^{*0}(5912)$ and $\Lambda_b^{*0}(5920)$ though their exact assignment as the $P$-wave $\Lambda_b$ has not been made. For simplicity, $\Lambda_{b}^{0}(5912)$ and $\Lambda_{b}^{0}(5920)$ are set to $\Lambda_b({1\over 2}^-)$ and $\Lambda_b({3\over 2}^-)$, respectively.

Based on numerical results, $\Sigma_b(6097)^\pm$ are very possibly $\Sigma_{b2}^{1}(\frac{3}{2}^-)$ or $\Sigma_{b2}^{1}(\frac{5}{2}^-)$ where there is no $\rho$-mode excitation inside. Under theoretical uncertainties, the total decay widths ($\Gamma\approx 19-20$ MeV ) are consistent with the experimentally measured ones by LHCb. In both assignments, $\Lambda_b^0\pi^\pm$ are their dominant decay channels with branching fraction ratios $\approx 72-74\%$.

The decay widths into $\Sigma_b^\pm\pi^0$ or $\Sigma_b^{*\pm}\pi^0$ are small in $\Sigma_{b2}^{1}(\frac{3}{2}^-)$ or $\Sigma_{b2}^{1}(\frac{5}{2}^-)$ assignment. In particular, the branching fraction ratio $R={\Gamma(\Sigma_b(6097)^\pm\to \Sigma_b^\pm\pi^0)\over \Gamma(\Sigma_b(6097)^\pm\to \Sigma_b^{*\pm}\pi^0)}$ is largely different in these two different assignments. If $\Sigma_b(6097)^\pm$ are $\Sigma_{b2}^{1}(\frac{3}{2}^-)$, the branching fraction ratios $R={\Gamma(\Sigma_b(6097)^\pm\to \Sigma_b^\pm\pi^0)\over \Gamma(\Sigma_b(6097)^-\to \Sigma_b^{*-}\pi^0)}=1.7$; If $\Sigma_b(6097)^\pm$ are $\Sigma_{b2}^{1}(\frac{5}{2}^-)$, the branching fraction ratios $R={\Gamma_{\Sigma_b(6097)^-\to \Sigma_b^-\pi^0}\over \Gamma_{\Sigma_b(6097)^-\to \Sigma_b^{*-}\pi^0}}=0.5$.

These ratios can be employed by experiment to distinguish $\Sigma_{b2}^{1}(\frac{3}{2}^-)$ from $\Sigma_{b2}^{1}(\frac{5}{2}^-)$. They depend weakly on the parameters in the $^3P_0$ model.

\section{Conclusions and discussions\label{Sec: summary}}
There are two ground $S$-wave and seven excited $P$-wave $\Sigma_b$ baryons. The OZI-allowed strong decay channels of these $\Sigma_b$ baryons have been given, and their widths have been systematically computed in the $^3P_0$ model.

$\Lambda_b^0\pi^\pm$ is the only strong decay channel of $\Sigma_b^\pm$ and $\Sigma_b^{*\pm}$. In comparison with experimental data, our theoretical results support the assignments that $\Sigma_b^\pm$ and $\Sigma_b^{*\pm}$ are the ground $S$-wave $\Sigma_b$ baryons with $J^P={1\over 2}^+$ and $J^P={3\over 2}^+$, respectively.

Channels $\Lambda_{b}^{0}\pi^\pm$, $\Sigma_{b}^\pm\pi^0$, $\Sigma_{b}^{*\pm}\pi^0$, $\Lambda_{b}^{0}(5912)\pi^\pm$ and $\Lambda_{b}^{0}(5920)\pi^\pm$ are five possible strong decays for $\Sigma_b(6097)^\pm$. Numerical results indicate that $\Sigma_b(6097)^\pm$ are very possibly $\Sigma_{b2}^{1}(\frac{3}{2}^-)$ or $\Sigma_{b2}^{1}(\frac{5}{2}^-)$. In these assignments, the decay widths $\Gamma\approx 19-20$ MeV, which are consistent with experimental measurements under theoretical uncertainties. $\Lambda_b^0\pi^\pm$ are their dominant decay channels with branching fraction ratios $\approx 72-74\%$.

For $\Sigma_{b2}^{1}(\frac{3}{2}^-)$ or $\Sigma_{b2}^{1}(\frac{5}{2}^-)$ assignment of $\Sigma_b(6097)^\pm$,  the ratios $R={\Gamma(\Sigma_b(6097)^\pm\to \Sigma_b^\pm\pi^0)\over \Gamma(\Sigma_b(6097)^\pm\to \Sigma_b^{*\pm}\pi^0)}$ are different. These ratios can be employed to distinguished $\Sigma_{b2}^{1}(\frac{3}{2}^-)$ from $\Sigma_{b2}^{1}(\frac{5}{2}^-)$, and they are expected to be measured in the future.

There are some uncertainties in the $^3P_0$ model. In addition to the masses of the hadron involved in the decay, the strong decay widths depend on some parameters such as $\gamma$ and $\beta$. In principle, $\beta$ can be derived directly in quark model. Unfortunately, $\beta$ of baryons have not yet been determined for a complexity of quark dynamics in baryons. $\beta$ were also set to those for mesons as in existed references. These uncertainties may change some predicted decay widths.

\begin{acknowledgments}
This work is supported by National Natural Science Foundation of China under the grant 11475111. Pei Yang and Jing-Jing Guo thank Dr. Ze Zhao very much for a helpful discussion.
\end{acknowledgments}


\begin{thebibliography}{99}
\bibitem{cdf}
T. Aaltonen {\it et al.} (CDF collaboration), Phys. Rev. Lett. {\bf 99}, 202001 (2007).
\bibitem{cdf2}
T. Aaltonen {\it et al.} (CDF collaboration), Phys. Rev. {\bf D 85}, 092011 (2012).
\bibitem{pdg}
M. Tanabashi {\it et al.} (Particle Data Group), Phys. Rev. D {\bf98}, 030001 (2018).
\bibitem{lhcb}
R. Aaij {\it et al.} (LHCb collaboration), arXiv: 1809.07752.
\bibitem{capstick}
S. Capstick and W. Roberts, Prog. Part. Nucl. Phys. 45, S241 (2000).
\bibitem{roberts}
W. Roberts and M. Pervin, Int. J. Mod. Phys. {\bf A 23}, 2817(2008).
\bibitem{klempt}
E. Klempt and Jean-Marc Richard, Rev. Mod. Phys. {\bf 82}, 1095 (2010).
\bibitem{crede}
V. Crede and W. Roberts, Rep. Prog. Phys. {\bf 76}, 076301 (2013).
\bibitem{cheng}
Hai-Yang Cheng, Front. Phys. 10, 101406 (2015).
\bibitem{chen1}
Bing Chen and Xiang Liu, arXiv: 1810.00389.
\bibitem{zhong1}
Kai-Lei Wang, Qi-Fang Lu and Xian-Hui Zhong, arXiv: 1810.02205.
\bibitem{micu1969}
L. Micu, Nucl. Phys. B {\bf 10}, 521 (1969).

\bibitem{yaouanc1}
A. Le Yaouanc, L. Oliver, O. P$\grave{e}$ne and J.C. Raynal, Phys. Rev. {\bf D 8}, 2223 (1973); {\bf 9}, 1415 (1974); {\bf 11}, 1272 (1975).
\bibitem{yaouanc2}
A. Le Yaouanc, L. Oliver, O. P$\grave{e}$ne and J.C. Raynal, Phys. Lett. {\bf B 71}, 397 (1977); {\bf B 72}, 57 (1977).
\bibitem{yaouanc3}
A. Le Yaouanc, L. Oliver, O. P$\grave{e}$ne and J.C. Raynal, {\it Hadron Transitions in the Quark Model} (Gordon and Breach Science Publishers, New York, 1987).
\bibitem{swanson}
P. Geiger, and E.S. Swanson, Phys. Rev. {\bf D 50}, 6855 (1994).
\bibitem{ackleh}
E.S. Ackleh, T. Barnes, and E.S. Swanson, Phys. Rev. {\bf D 54}, 6811 (1996).
\bibitem{bonnaz}
R. Bonnaz, L.A. Blancob, B. Silvestre-Brac, F. Fern¨¢ndez, A. Valcarce, Nucl. Phys. {\bf A 683}, 425 (2001).
\bibitem{zhu}
Chong Chen, Xiao-Lin Chen, Xiang Liu, Wei-Zhen Deng and Shi-Lin Zhu, Phys. Rev. {\bf D 75}, 094017 (2007).
\bibitem{zhang}
Ze Zhao, Dan-Dan Ye, and Ailin Zhang, Phys. Rev. {\bf D 94}, 114020 (2016).
\bibitem{zhang2}
Ze Zhao, Dan-Dan Ye, and Ailin Zhang, Phys. Rev. {\bf D 95}, 114024 (2017).
\bibitem{zhang3}
Dan-Dan Ye, Ze Zhao, and Ailin Zhang, Phys. Rev. {\bf D 96}, 114003 (2017).
\bibitem{zhang4}
Dan-Dan Ye, Ze Zhao, and Ailin Zhang,  Phys. Rev. {\bf D 96}, 114009 (2017).
\bibitem{capstick2}
S. Capstick and N. Isgur, Phys. Rev. {\bf D 34}, 2809 (1986).
\bibitem{roberts2}
S. Capstick and W. Roberts, Phys. Rev. {\bf D 47}, 1994 (1993).
\bibitem{jacobi}
C. G. J. Jacobi, {\it Vorlesungen $\ddot{u}$ber Dynamik, Gesammelte Werke}, Vol. VIII, Supplement (1884); reprinted by (Chelsea Publishing Company, New York, 1969).
\bibitem{cheng2}
Hai-Yang Cheng and Chun-Khiang Chua, Phys. Rev. {\bf D 75}, 014006 (2007); Phys. Rev. {\bf D 92}, 074014 (2015).
\bibitem{hosaka}
T. Yoshida,E. Hiyama, A. Hosaka, M. Oka and K. Sadato, Phys. Rev. {\bf D 92}, 114029 (2015).
\bibitem{ebert}
D. Ebert, R.N. Faustova and V.O. Galkin, Phys. Lett. {\bf B 659}, 612 (2008); Phys. Rev. {\bf D 84}, 014025 (2011).
\bibitem{rosner}
M. Karliner and J.L. Rosner, Phys. Rev. {\bf D 92}, 074026 (2015).
\bibitem{rai}
K. Thakkar, Z. Shah, A.K. Rai, and P. C. Vinodkumar, Nucl. Phys. {\bf A 965}, 57 (2017).
\bibitem{godfrey}
H.G. Blundell and S. Godfrey, Phys. Rev. {\bf D 53}, 3700 (1996).
\bibitem{godfrey3}
S. Godfrey, K. Moats, and E. S. Swanson, Phys. Rev. {\bf D 94}, 054025 (2016).
\bibitem{godfrey2}
S. Godfrey and K. Moats, Phys. Rev. {\bf D 93}, 034035 (2016).
\bibitem{lhcb2}
R. Aaij {\it et al.} (LHCb collaboration), Phys. Rev. Lett. {\bf 109}, 172003 (2012).
\end{thebibliography}
\end{document}